\documentclass{sig-alternate-2013}

\newfont{\mycrnotice}{ptmr8t at 7pt}
\newfont{\myconfname}{ptmri8t at 7pt}
\permission{}
\conferenceinfo{}{} 
\copyrightetc{}
\crdata{}
\usepackage[T1]{fontenc}
\usepackage{float}
\usepackage{graphicx}
\usepackage{array}
\usepackage{url}
\usepackage{cite}

\newcommand{\hide}[1]{}

\begin{document}
\title{Evaluating the Effect of Centralization on Routing Convergence on a Hybrid BGP-SDN Emulation Framework}

\numberofauthors{3}
\author{
\alignauthor{Adrian G\"amperli} \\
\affaddr{ETH Zurich}\\
\affaddr{Zurich, Switzerland}\\
\email{gaadrian@student.ethz.ch} 
\alignauthor{Vasileios Kotronis}\\
\affaddr{ETH Zurich}\\
\affaddr{Zurich, Switzerland}\\
\email{vkotroni@tik.ee.ethz.ch}
\alignauthor{Xenofontas Dimitropoulos}\\
\affaddr{FORTH}\\
\affaddr{Heraklion, Greece}\\
\email{fontas@ics.forth.gr}
}

\maketitle

\section{Motivation}
A lot of applications depend on reliable and stable Internet connectivity. 
These characteristics are crucial for mission-critical
services such as telemedical applications. An important factor
that can affect connection availability is the convergence time of BGP,
the de-facto inter-domain routing (IDR) protocol in the Internet.
After a routing change, it may take several minutes until the network 
converges and BGP routing becomes stable again~\cite{quantifyingoliveira}.  
Kotronis \emph{et al}.~\cite{outsourcingkotronis,ONS-RaaS} propose a novel Internet routing
approach based on SDN principles that combines several Autonomous Systems (AS) into groups, 
called clusters, and introduces a logically centralized routing decision process 
for the cluster participants. 
One of the goals of this concept is to stabilize the IDR system
and bring down its convergence time.
However, testing whether such approaches can improve on BGP problems
requires hybrid SDN and BGP experimentation tools that can emulate
multiple ASes.
Presently, there is a lack of an easy to use public tool for this
purpose. This work fills this gap by building a suitable emulation
framework and evaluating the effect that a proof-of-concept IDR
controller has on IDR convergence time.

\\

\section{Objectives}
Our primary objective is to support
hybrid BGP-SDN experiments with multiple ASes using real router software. 
This is needed since when deploying a new IDR approach one cannot 
change the whole infrastructure at once. 
The framework should take care of configuration management such as IP prefixes and BGP policy
templates and the user should be able to actively control the experiments, e.g., dynamically 
changing the topology and verifying the effects of changes. Furthermore, it should be 
possible to easily create topologies based on measured Internet data or theoretical models. 
This way an experimenter should be able to concentrate more on the experiments and her concept
rather than bothering with configuration and management.

Our second objective is to demonstrate the effect of centralization on
IDR convergence time. We designed and implemented a proof-of-concept IDR SDN
controller that exploits centralization to improve IDR
convergence time based on the following design goals. First, the controller
should inter-operate with legacy BGP routers. Moreover, the cluster
network is transparent to the legacy BGP world, therefore ASes within
the cluster maintain their AS identity. In addition, we want to support
disjoint AS sub-clusters controlled by the same controller, so that an
intra-cluster link failure does not isolate the controlled ASes: paths
over the legacy Internet could still connect the sub-clusters. 

\section{Hybrid SDN \& BGP Emulation \\Framework}

The framework is based on a slightly modified version of Mininet~\cite{mininet} that
supports the functionality of Quagga~\cite{quagga} -- a popular BGP software.
The topologies can be either artificial or built from the iPlane Inter-PoP links~\cite{iplane} 
and the CAIDA AS Relationship~\cite{caida} datasets. 
The framework automatically assigns IP addresses and configures
network devices, including customer-to-provider and peer-to-peer
relationships.
To isolate the effects of inter-domain from intra-domain routing every AS 
is emulated by a single network device. This abstraction is useful for use-cases
such as ours.

\begin{figure}[h!]
\begin{centering}
\includegraphics[width=0.4\textwidth]{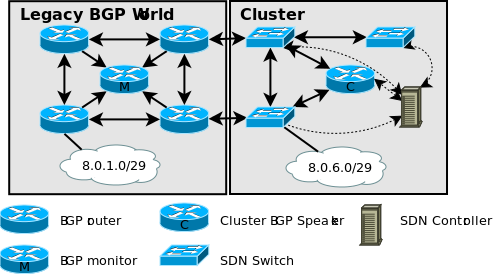}
\caption{The different components of an example hybrid BGP/SDN
  emulation experiment.}
\label{fig:frameworkoverview}
\end{centering}
\end{figure}

In Fig.~\ref{fig:frameworkoverview} we show the components of an
example experiment. On the left side, we see the legacy BGP part of
the emulated network, whereas on the right side we illustrate an SDN
cluster, composed of OpenFlow switches. BGP routers and SDN switches
can originate prefixes. It is also possible to add hosts with IP
addresses within a particular prefix for monitoring end-to-end
connectivity with tools like ping, etc. All BGP routers peer with a
BGP route collector, which collects routing updates for monitoring
purposes. Moreover, within the SDN cluster we have a special BGP speaker, called
{\it cluster BGP speaker}, which relays routing information between
external BGP routers and the SDN controller. The cluster BGP speaker
is implemented with ExaBGP~\cite{exabgp}. For every BGP peering there
is a link from the cluster BGP speaker to the border SDN switch, so
as to relay control plane information over the switches.

Experimental setups can be written in Python. We implemented several
additional Mininet-BGP commands to announce prefixes, wait
until BGP has converged, etc. Additionally, the framework supports
tools for automatic log file analysis, network graph creation,
convergence time and loss measurement, and route change
visualization. For example, to facilitate experiments on IDR
stability, the framework detects when the network has
converged and whether there is stable connectivity between
all hosts. Other compatible tools can be added as Mininet is an
extensible platform.

We built an IDR SDN controller over the cluster BGP speaker (using
POX~\cite{pox}), to evaluate the effect of IDR centralization on
convergence time. More details on the design and implementation of the
IDR controller can be found at~\cite{mt}. An important insight that we
gained is that we can not naively use the same loop avoidance
mechanism as BGP, 
due to the differences between the distributed
path selection of BGP and the centralized routing control of SDN. We
therefore introduce two graphs for the route selection process: the
\emph{Switch graph}, representing the physical topology of the
switches in the cluster and the \emph{AS topology graph}, which is a
transformation of the switch graph per destination prefix. The
transformation is restructuring the graph taking carefully into account
paths that cross the legacy world and the SDN cluster so as to avoid
loops. Best path calculations are based on the Dijkstra algorithm,
running on the AS topology graph. AS routes are then compiled to flow
rules on the SDN switches. Another design insight we gained is the
need for a delayed recomputation of best paths on the controller's
side, so as to improve overall stability and rate-limit route flaps
due to bursts in external BGP input.

\section{Demo and Results}

\begin{figure}[h!]
\centering
\includegraphics[trim = 0mm 0mm 0mm 20mm ,width=0.35\textwidth]{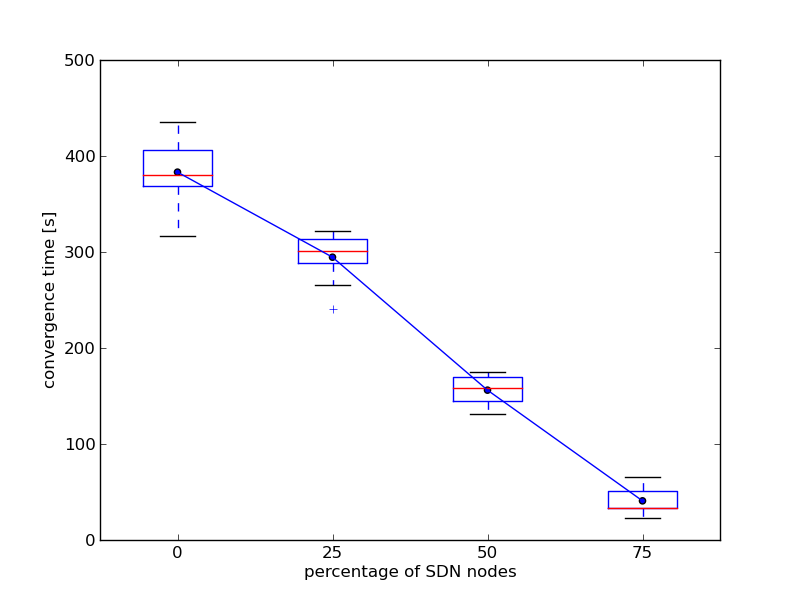}
\caption{IDR convergence time of route withdrawal on a 16-AS clique
  topology versus fraction of ASes with centralized route control. The
  remaining ASes use standard BGP. We show boxplots over 10 runs.}
\label{fig:withdrawal}
\end{figure}

The demonstration will show the framework and our use-case. One part of
the demo will focus on how researchers can use the framework to run
and manage experiments~\cite{demovideo}.
Secondly, we will demonstrate the effect of
SDN centralization on IDR convergence time showing visually how it
affects an end-to-end video application under different scenarios. Our
experiments on a clique topology~\cite{mt} show that IDR
centralization can improve the convergence time even with small SDN
cluster deployments. In Fig.~\ref{fig:withdrawal} we show how the
convergence time can be linearly reduced in a route withdrawal
experiment with different percentages of SDN deployment in a 16-node
clique. On the other hand, route fail-over and announcement
experiments did not show this linear improvement, but smaller
reductions.

\section{Related Work}

MiniNext~\cite{schlinker2014try} is a hybrid SDN - legacy routing
emulator based on Mininet and Quagga. However, while MiniNext aims at
emulating operational environments and focuses on low-level APIs, our
framework focuses on multi-AS IDR experiments and provides a
high-level API for experiment lifecycle orchestration. 
The Open Network Operating System (ONOS)~\cite{onos} is built to be production
ready for large-scale networks. Our approach is better suited for rapid
prototyping. Due to simplifications such as cooperative multitasking, 
we can focus more on research questions than on
state consistency and concurrency issues.  Finally,
RouteFlow~\cite{routeflow} is a platform where the controller
application mirrors the SDN topology to a virtual network and runs a
legacy routing protocol on top of it. 
Our controller however does not rely on routing decisions of legacy
protocols but
runs its own algorithms, enabling better integration with SDN concepts.

{\bf Acknowledgements:}~This work was partly funded by European Research Council Grant Agreement n. 338402.

\bibliographystyle{abbrv}
{\scriptsize
\setlength{\itemsep}{0pt}
\bibliography{pd148-gaemperli}

\begin{thebibliography}{10}

\bibitem{demovideo}
\url{http://youtu.be/Cbc8XlIp_C0}.

\bibitem{pox}
{POX}.
\newblock \url{http://www.noxrepo.org/pox/about-pox/}.

\bibitem{quagga}
{Quagga Routing Software Suite}.
\newblock \url{http://www.nongnu.org/quagga/}.

\bibitem{schlinker2014try}
{B. Schlinker et al.}
\newblock {Try Before you Buy: SDN Emulation with (Real) Interdomain Routing}.
\newblock In {\em Proc. of ONS}, 2014.

\bibitem{routeflow}
{C. E. Rothenberg et al.}
\newblock {Revisiting Routing Control Platforms with the Eyes and Muscles of
  Software-defined Networking}.
\newblock In {\em Proc. of ACM HotSDN}, 2012.

\bibitem{exabgp}
{Exa Networks}.
\newblock {Exa-Networks/exabgp}.
\newblock \url{https://github.com/Exa-Networks/exabgp}.

\bibitem{mt}
A.~G\"amperli.
\newblock {Evaluating the Effect of SDN Centralization on Internet Routing
  Convergence}.
\newblock Master's thesis, ETH Z\"urich, 2014.

\bibitem{outsourcingkotronis}
V.~Kotronis, X.~Dimitropoulos, and B.~Ager.
\newblock {Outsourcing the Routing Control Logic: Better Internet Routing Based
  on SDN Principles}.
\newblock In {\em Proc. of ACM HotNets-XI}, 2012.

\bibitem{ONS-RaaS}
V.~Kotronis, X.~Dimitropoulos, and B.~Ager.
\newblock {Outsourcing Routing using SDN: The Case for a Multi-Domain Routing
  Operating System}.
\newblock In {\em Poster Proc. of ONS}, 2013.

\bibitem{mininet}
B.~Lantz, B.~Heller, and N.~McKeown.
\newblock {A Network in a Laptop: Rapid Prototyping for Software-defined
  Networks}.
\newblock In {\em Proc. of ACM HotNets-IX}, 2010.

\bibitem{caida}
\mbox{CAIDA}.
\newblock As relationships dataset.
\newblock \url{http://www.caida.org/data/as-relationships/}.

\bibitem{iplane}
\mbox{University of Washington}.
\newblock {iPlane: Datasets}.
\newblock \url{http://iplane.cs.washington.edu/data/data.html}.

\bibitem{quantifyingoliveira}
{Oliveira, Ricardo et al.}
\newblock {Quantifying Path Exploration in the Internet}.
\newblock In {\em Proc. of ACM IMC}, 2006.

\bibitem{onos}
ON.Lab.
\newblock {What is ONOS?}
\newblock \url{http://tools.onlab.us/onos.html}.

\end{thebibliography}
}

\end{document}